\documentclass[]{spie}  

\usepackage{amsmath,amsfonts,amssymb}
\usepackage{graphicx}
\usepackage[colorlinks=true, allcolors=blue]{hyperref}
\usepackage{multirow}  
\usepackage{eurosym}
\usepackage[colorlinks=true, allcolors=blue]{hyperref}

\title{Operational forecast of the PSF figures of merit}

\author[a]{Turchi, A.}
\author[a]{Agapito, G.}
\author[a]{Masciadri, E.}
\author[b,c]{Beltramo-Martin, O.}
\author[a]{Pinna, E.}
\author[b,c]{Sauvage, J.F.}
\author[b,c]{Fusco, T.}
\author[b]{Neichel, B.}
\author[d]{Milli, J.}
\affil[a]{INAF-Osservatorio Astrofisico di Arcetri, L.go Enrico Fermi 5, Firenze, Italy}
\affil[b]{LAM-Laboratoire d'Astrophysique de Marseille, UMR 7326, 13388, Marseille, France}
\affil[c]{ONERA, B.P. 72, F-92322 Chatillon, France}
\affil[d]{IPAG-Institut de Planétologie et d’Astrophysique de Grenoble, 414, Rue de la Piscine, 38400 St-Martin d’Hères, France}

\authorinfo{Send correspondence to Alessio Turchi - e-mail: alessio.turchi@inaf.it}

\begin{document}
\maketitle
\begin{abstract}
The optimization and scheduling of scientific observations done with instrumentation supported by adaptive optics could greatly benefit from the forecast of PSF figures of merit (FWHM, Strehl Ratio, Encircle Energy and contrast), that depend on the AO instrument, the scientific target and turbulence conditions during the observing night. In this contribution we explore the the possibility to forecast a few among the most useful PSF figures of merit (SR and FWHM). To achieve this goal, we use the optical turbulence forecasted by the mesoscale atmospheric model Astro-Meso-NH on a short timescale as an input for PSF simulation software developed and tailored for specific AO instruments. A preliminary validation will be performed by comparing the results with on-sky measured PSF figures of merit obtained on specific targets using the SCAO systems SOUL (FLAO upgrade) feeding the camera LUCI at LBT and SAXO, the extreme SCAO system feeding the high resolution SPHERE instrument at VLT. This study will pave the way to the implementation of an operational forecasts of such a figure of merits on the base of existing operational forecast system of the atmosphere (turbulence and atmospheric parameters). In this contribution we focus our attention on the forecast of the PSF on-axis.
\end{abstract}


\section{INTRODUCTION}
\label{sec:intro}  

So far the abilities of models (atmospheric models, statistical models, etc.) in predicting the optical turbulence (OT) have been investigated. We know today that we can predict the optical turbulence (OT) with a very good accuracy,  specifically the most relevant astroclimatic parameters, and we are able to do that in an operational configuration [\cite{masciadri2020}]. The next step is therefore to ask to ourselves if we can directly predict the point spread function (PSF) figures of merits. Specifically we are interested in forecasting the full width at half maximum (FWHM), the Strehl ratio (SR), the encircled energy (EE) and the contrast.

In ground-based observations supported by the adaptive optics (AO) the estimation of the PSF figures of merit is not a trivial problem. The complexity of such a topic is due to the fact that the PSF depends on the atmospheric conditions, on the physical characteristics of the observed objects and also on the instrument specifications. If we want to forecast (and not simply measure) these figures of merit the goal is even more challenging.
Besides, the estimate of the PSF on-axis and off-axis implies a different level of complexity. If the PSF on-axis allows a simpler analysis as it can be treated as a scalar problem, because the figures of merits mainly depends on the integral of the optical turbulence along the line of sight (seeing $\epsilon$ or Fried parameter $r_0$), it is known that the PSF off-axis is a more complex problem as it depends on the vertical stratification of the turbulence in the atmosphere. This is due to the fact the direction with respect to which perturbations affecting the wavefront are measured is different from that along which the scientific object is observed. The turbulence that the wavefront passes through at the different heights of the atmosphere along the different lines of sight can therefore be different and the result is a non-optimal wavefront correction, that causes a distortion and an elongation of the PSF at large angular distances from the reference stars, the so called `anisoplanatism effect' [\cite{tallon1990}]. More sophisticated AO systems are used in order to overcome this problem, such as the laser-tomographic AO (LTAO), the multi-conjugated AO (MCAO) or the multi-object AO (MOAO) [\cite{rigaut2018}], whose main goal is to correct the wavefront perturbations on a field of view of a few arcminutes (typically 1-2 arcminutes for the LTAO, 2-3 arcminutes for the MCAO and 5-10 arcminutes for the MOAO). Such systems allows a more uniform and homogenous PSF on a large FOV but they require the knowledge of the vertical stratification of the optical turbulence ($C_{N}^{2}$) and the wind speed. Both of them can be forecasted with the models/methods cited above [\cite{masciadri2013,masciadri2017,masciadri2020}].

The idea behind this contribution is to present a roadmap of a long-term study whose final goal is that to forecast the PSF figures of merits in operational configuration to be applied in different contexts, i.e. observations supported by narrow field of view (FOV) AO systems (SCAO systems) and observations supported by wide field of view adaptive optics (WFAO) where the problem of the off-axis PSF becomes relevant.

To simplify our approach and reduce the number of free parameters, we start analysing the PSF on-axis. For the forecast of the atmosphere we choose a hybrid system given by the simultaneous use of the Astro-Meso-Nh model and the autoregressive method (AR) that are able to forecast the optical turbulence with a very good accuracy [\cite{masciadri2020}]. The forecasts are used as an imput for the AO modeling software that is able to reconstruct the operation of an AO system from the passage of a wavefront through the atmosphere up the formation of the image of the detector. There are two typologies of such a models: the numerical end-to-end models and the analytical models. Both of them have pros and cons. We choose in this first step, an end-to-end model (PASSATA) [\cite{agapito2016}] that is able to simulate a PSF and retrieve in post-processing various PSF figures of merit. We limit our attention, in this contribution, to the SR and FWHM but other might be implemented (e.g. EE, contrast, etc).  PASSATA has to be configured for the two AO systems: SOUL (at LBT) and SAXO (at the VLT). We will analyse in this contribution the case of SOUL [\cite{pinna2019}]. The configuration of PASSATA using technical specification for SAXO is in progress.

The final goal of this study is to set-up a system that is able to provide a forecast of the PSF figures of merit using an approach similar to the one used for the optical turbulence by ALTA Center. As described in Section \ref{model}, ALTA provides a long time scale forecast (the afternoon before the coming night) and a short time scale forecast that starts to produce data during the course of observations. The second approach provides, at each full hour of the night, a more accurate forecast extended in the successive four hours. Our intention is to perform a similar procedure for the SR and the FWHM. When the night starts, the idea is to provide, at each full hour, a forecast on the same time scale for three different magnitudes covering the whole range of magnitudes typically used by the SOUL and SAXO. In general we are interested in supporting the flexible scheduling of AO-assisted observations and in improving the efficiency of AO instruments.

In Section \ref{model} we introduce the ALTA atmospheric model setup used for this study. In Section \ref{passata} we describe the AO modeling software configuration and the relevant key points related to this part of the implementation. In Section \ref{config} we report the preliminary configuration used in this study in order to produce the PSF forecasts of FWHM and SR with the joined atmospheric and AO modeling systems. In Section \ref{results} we present the results obtained with the FWHM and SR forecasts by comparing with the measurements coming from the SOUL AO system. Finally in the conclusion we sum up the results of this work,

\section{THE ATMOSPHERIC MODEL}
\label{model}

Meso-NH [\cite{lafore98,lac2018}] is a mesoscale atmosphere model that allows simulating the temporal evolution of atmospheric parameters in a volume defined over a finite geographical grid. Together with Astro-Meso-NH code [\cite{masciadri1999}], it is also able to model the OT parameters including the full $C_N^2$ stratification.  We refer to Turchi et al. [\cite{turchi2017}] for more details of the numerical configuration used for the ALTA Center project. ALTA uses initial conditions (used for model initialization and periodical forcing of external conditions) provided by the General Circulation Model (GCM) run by the European Center for Medium Weather Forecasts (ECMWF), with an horizontal resolution of 9 km. ALTA Center simulates the whole observing night (approximately between 00:00 UT and 15:00 UT) at Mount Graham, where the LBT site is located (coordinates [32.70131, -109.88906], at an height of 3221 m above sea level).
The standard forecast produced by the model is provided during the afternoon before the start of the night, however in recent years this system has been upgraded with a short-term forecast based on the usage of autoregression (AR) techniques [\cite{masciadri2020}]. This method uses the real-time measurements produced by the telescope telemetry (atmospheric sensors, DIMM monitor, etc...) in order to learn from the forecast discrepancies over the previous nights and improve the accuracy of the predictions over the next few hours in the future (currently up to 4 hours). This allows to maximize the accuracy and still gives precious information for observation planning.. This short-term forecast is produced each hour during the night and allows extreme forecast accuracy [\cite{masciadri2020}], up to a RMSE of 0.1 arcsec with respect to the seeing parameter at an hour in the future [\cite{masciadri2020}].\\
A system similar to ALTA is currently under development for ESO at VLT\footnote{The operational forecast system for the VLT is not an official ESO tool.} and it has already been possible to prove its abilities in providing PWV forecast at very high accuracy [\cite{turchi2020}] at short time scales.

In this contribution we are interested in trying to assess the best possible performance obtainable for the PSF forecast in order to drive the specifications of the next phases in which the real operational system will be constructed. For this reason, in this study we will use the ALTA short-term forecast as an input to PASSATA AO instrument simulations in order to obtain the best possible PSF forecast.

\begin{figure}
\centering
\includegraphics[width=.6\textwidth]{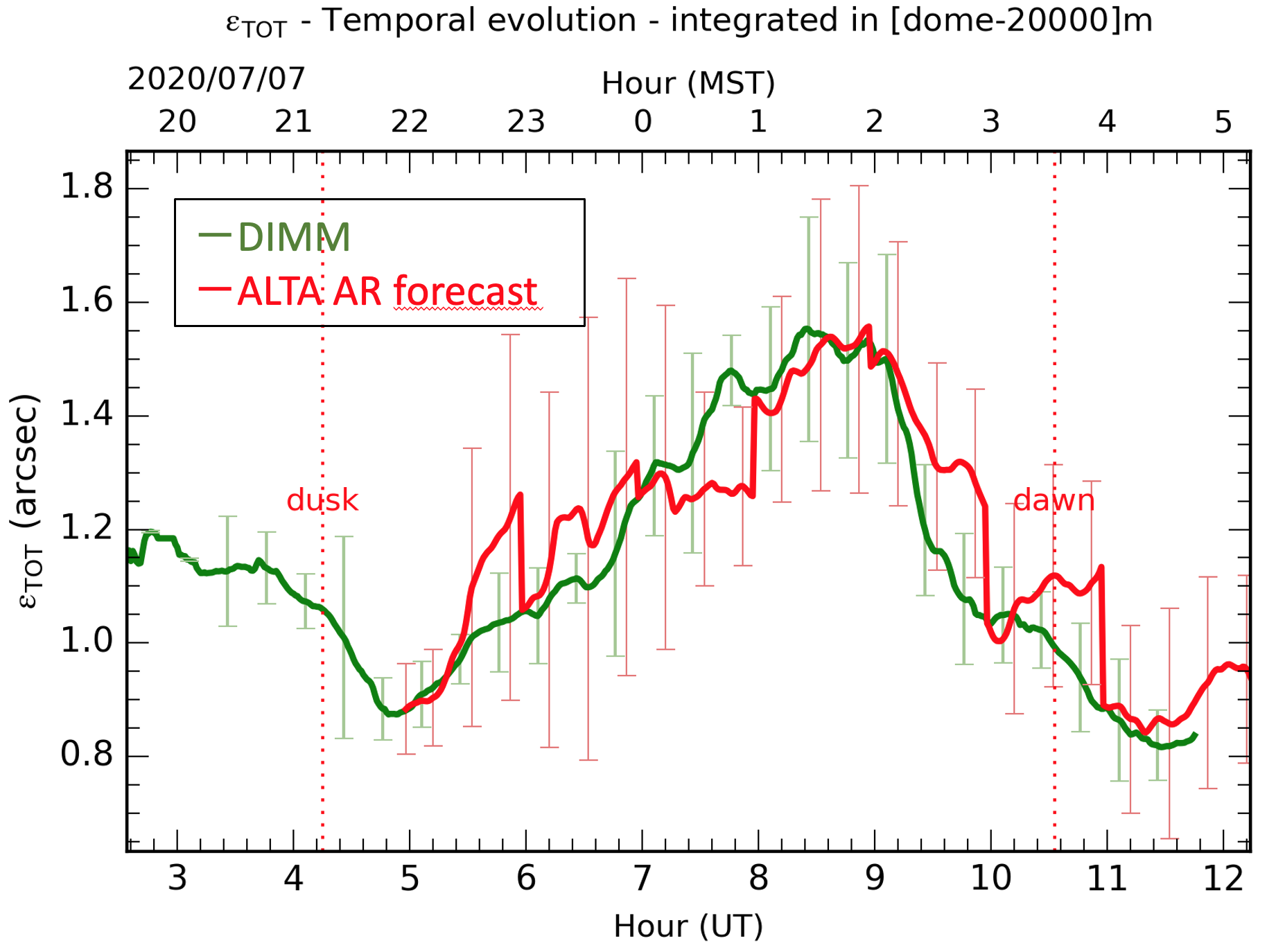}
\caption{Example of a forecast of the seeing obtained with the AR filter at a time scale of 1h.}
\label{fig:AltaFor}
\end{figure}

\section{THE AO MODELING SOFTWARE}
\label{passata}

The outputs of the ALTA system, describing the status of the OT and the atmosphere, are injected into an end-to-end AO numerical modeling software (PASSATA[\cite{agapito2016}]) that is able to reconstruct the operation of an AO system from the passage of a wavefront through the atmosphere up to the formation of the image of the detector. Among the PASSATA outputs we mention therefore also the PSF and its figures of merit calculated in a post-processing phase. PASSATA can be adapted to any specific AO system by including their technical specifications.  In this contribution we limit our attention to the Strehl ratio (SR) and the full width at half maximum (FWHM) but other might be implemented (e.g. Encircle Energy, contrast, etc). In this first contribution we limit our analysis to the SOUL AO system used at the Large Binocular Telescope (LBT). The setting-up of PASSATA with the SAXO specifications (AO system of SPHERE at the VLT) is  currently in rapid progress.

It is known that the end-to-end models are in principle more complete and accurate than analytical ones but they require longer computational time. We therefore implemented a GPU dedicated server to be able to perform the simulation within a few minutes. This condition is necessary as our aim is to perform an operational short-term forecast of the PSF figures of merit in order to use the outputs of such a forecasts for the scheduling of the scientific programs, during the observing night.

\begin{figure}
\centering
\includegraphics[width=1.\textwidth]{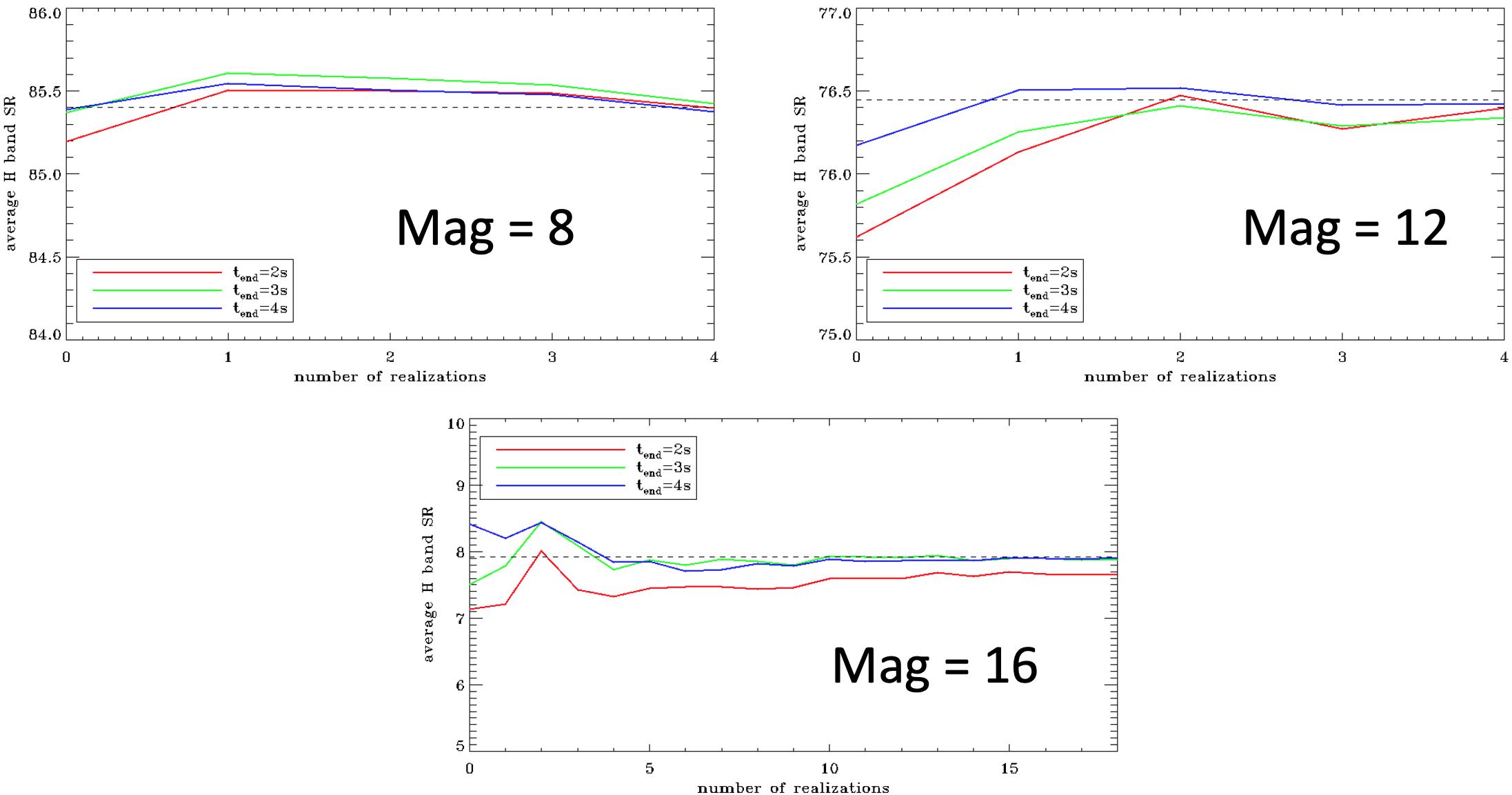}
\caption{PASSATA convergence times with respect to the number of realizations. Plots were obtained with magnitudes 8, 12 and 16. The different lines correspond to different lengths of the simulated time. The dashed line represents the median value which is the target of the convergence.}
\label{fig:converge}
\end{figure}

PASSATA is an IDL and CUDA based object oriented library/software capable of doing Monte-Carlo end-to-end Adaptive Optics (AO) simulations. GPU acceleration is optional and limited to the most computationally demanding routines. CUDA functions are wrapped in a DLM extension that can be loaded by IDL at runtime. The end-to-end simulations constitute the modelling of all the steps involving the AO system main functionality and the input disturbance generation and propagation from the guide source to the wavefront sensor through the atmosphere and the deformable mirror.
Inputs are the telescope, AO system and turbulence parameters and outputs are AO telemetry (slopes, mirror commands, ...) and residual phase.
In a second step, PSF is computed from the residual phase at the desired wavelength, and then, SR and FWHM are evaluated on this PSF. This process can be repeated to get performance for different wavelengths.

To perform simulations that are fast enough for an operational configuration we calculated the time necessary to be simulated for different magnitudes to converge on a stable solutions. This has been done averaging results from a set of simulations done with different atmospheric and noise realizations. Given the current specifications, we initially plan to deliver a forecast of FWHM and SR with a timestep of 20 minutes, in order to have at least 3 points each hour, and for only few sample magnitudes.  The selected configuration determines the number of GPUs required. The definitive setup is still to be defined but the idea behind it is that we want to guarantee a useful information but with limited costs. For this initial test we chose three magnitudes of 8, 12 and 16, which cover the typical range of observing targets.\\
We chose different simulation times ranging from 2 to 4 seconds. As expected when the magnitude is brighter, thus the correction is better, a small number of simulations are required to get close to the median value as can be seen in Fig. \ref{fig:converge}. In the end we chose to run 1 simulation of 2s for magnitude 8 case, 3 simulations of 2s for magnitude 12 and 5 simulations of 3s for the faintest magnitude (16).
With the previous setup we were able to obtain simulated outputs for the forecast in less than ten minutes for the different configurations, showing the feasibility of an almost-real-time operational forecast. With the above setup, we need at least one GPU for each simulation on the specified magnitudes and for each timestep. In order to cover few hours of forecast with at least three magnitudes in the near future, we are planning to either improve the performances by powering up our GPU hardware (additional GPUs or GPU servers) or either using a different faster (analytical) model, thus providing ample space for performance improvements.

\section{PRELIMINARY CONFIGURATION}
\label{config}

Several challenges lie ahead before being able to finalize the specifications for a PSF forecast tool. A key element is the quantification of the accuracy of the PSF measurements obtained by the AO instruments themselves, that is the reference with respect to which one has to quantify the goodness of a forecast. This means to define all the elements that contribute to the measured PSF variability. The other main challenge is to compare the accuracy of measurements with the performance of the forecasts, in the context of science operations, that is well different from the AO characterization during commissioning. A common source of uncertainty are, for example, telescope structure vibrations and non-common-path aberrations (NCPA).

As a preliminary step, in order to simplify our approach and reduce the number of free parameters, we start analysing the PSF on-axis for the SOUL case. PASSATA has been used extensively in SOUL design [\cite{pinna2016}], however we had to modify the configuration in order to create a setup which is compatible with forecast requirements, as discussed in section \ref{passata}.\\
SOUL has a database of PSF measurements available, thus giving us the possibility to produce a preliminary evaluation of the feasibility of the PSF forecast, in terms of full width at half maximum (FWHM) and Strehl ratio (SR). Uncertainties on the final PSF forecast are coming both from the atmospheric model and from the AO modeling software. 

For the forecast of the atmosphere we choose a hybrid system given by the simultaneous use of the Astro-Meso-Nh model and the AR method that is able to forecast the optical turbulence with a very good accuracy [\cite{masciadri2020}].

The short term atmospheric and OT forecasts that are used as an input for PASSATA are provided by the operational forecast system named ALTA Center (\href{http://alta.arcetri.inaf.it}{http://alta.arcetri.inaf.it}) running at LBT.
In this study we selected, among the available set which had a corresponding PSF measurement provided by SOUL, the best matching seeing forecasts produced by the ALTA system with the short-term (AR) method. Specifically we selected the time windows in which the agreement between the AR forecast and the DIMM measurements was optimal. This will allow us to disentangle the analysis of contribution to the total uncertainty coming from the atmospheric model (Astro-Meso-Nh+AR) from those coming from the differences between the measured PSF (with SOUL) and the simulated PSF with the AO modeling software (PASSATA).

We limit ourselves to the PSF on-axis therefore we consider a fixed stratification of the optical turbulence and the wind speed. Specifically we consider both wind speed and $C_N^2$ distributed in four layers of infinitesimal width located at 119m, 837m, 3045m, and 12790m (distances are given with respect to the conjugated pupil plane). On each layer the wind speed is 2 m/s, 4 m/s, 6 m/s and 25 m/s respectively, while the $C_N^2$ is proportionally distributed on each layer with the following total percentages: 70\%, 6\%, 14\% and 10\%.\\
This is the configuration that is currently used for the calculation of the SOUL SR  during the commissioning and it refers to typical conditions. We decided to use those values as a first reference as it allow us to reduce the free parameters in this preliminary evaluation and it allows us to investigate the sources of uncertainties related to the AO modeling tool. Indeed since on-axis PSF mainly depends on the integrated quantities, the stratification of the $C_N^2$ has a second-order impact on the final results. In future developments, especially when moving towards off-axis forecasts, the variability of the $C_N^2$ will be taken into account.
In the present preliminary configuration we inject in PASSATA only the seeing coming from Astro-Meso-Nh + AR corrected by the airmass correspondent to the line of sight of the specific astronomical observations.

Beside the instrument and atmospheric OT specifications, PASSATA has to be configured with the observed target characteristics, specifically the magnitude of the reference star. Once this final ingredient is inserted, PASSATA produces as an output the correspondent PSF and its figures of merit (FWHM and SR).

We resume here the steps taken so far in developing this preliminary version of the PSF forecast tool, which is one of the goals of this preliminary work:
\begin{itemize}
    \item We installed a GPU dedicated server allowing us to perform simulations with the end-to-end PASSATA software in a reasonable short time scale (few minutes).
    \item We identified temporal windows in which the agreement of the observed and forecasted seeing was good for a few test cases as this allows us to study the uncertainties of the atmospheric module and the AO models.
    \item We perform the simulations with PASSATA using as inputs seeing (from ALTA+AR forecast corrected by airmass) and magnitude of the reference star.
    \item From PASSATA simulations, we computed the FWHM and the SR for the wavelengths corresponding to SOUL PSF measurements (H, 1.64 $\mu m$, and Ks, 2.16 $\mu m$, bands) and we compared the results between the two.\\
    \item At present, AO vibrations and NCPA are not removed from the dataset.
\end{itemize}

\section{RESULTS}
\label{results}

\begin{figure}
\centering
\includegraphics[width=1.\textwidth]{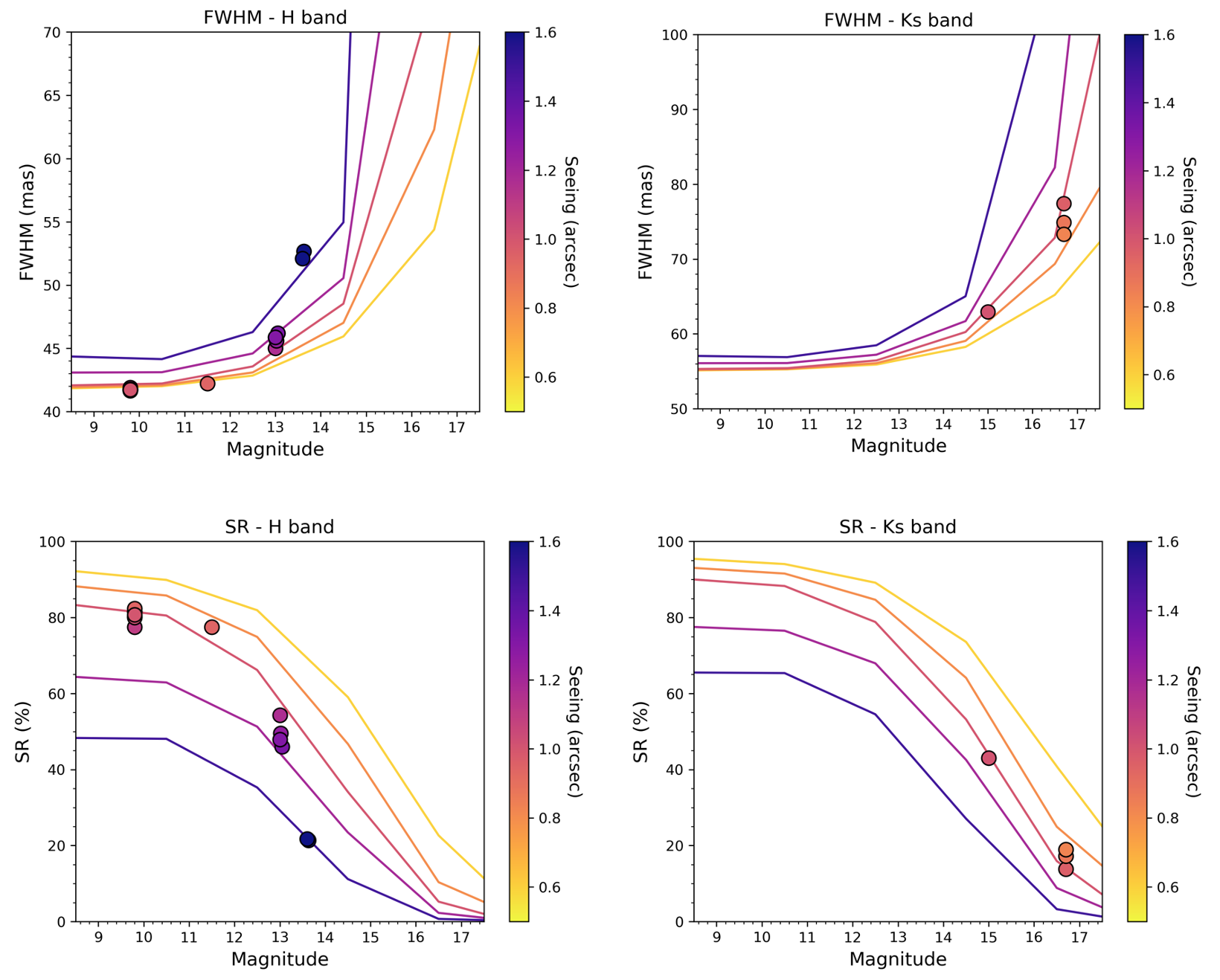}
\caption{Simulated curves for FWHM (upper row) and SR (lower row) in H (left) and Ks (right) bands respectively, with different magnitudes from 8.5 to 17.5, obtained with PASSATA as configured for SOUL. The dots represent the ALTA + PASSATA simulations from Table \ref{fig:TableRes} obtained with our forecast configuration.  In the colorbar we represent the different seeing values used for the simulations, ranging from 0.6 to 1.5 arcsec. Each data point represents a median of the respective quantity over a 10-minutes time interval.}
\label{fig:FigPassata}
\end{figure}

In this section we present the results obtained with the approach described before. In fig. \ref{fig:FigPassata} we report the PASSATA curves obtained with the standard SOUL configuration used for the commissioning evaluation, for different magnitudes and seeing values, for both FWHM and SR in H and Ks bands, together with the ALTA + PASSATA simulations that we performed with our modified PASSATA for the forecast configuration. The agreement shows the correctness of our PASSATA implementation with respect to the reference SOUL design.\\

\begin{table}
\label{fig:TableRes}
\centering
\includegraphics[width=1.\textwidth]{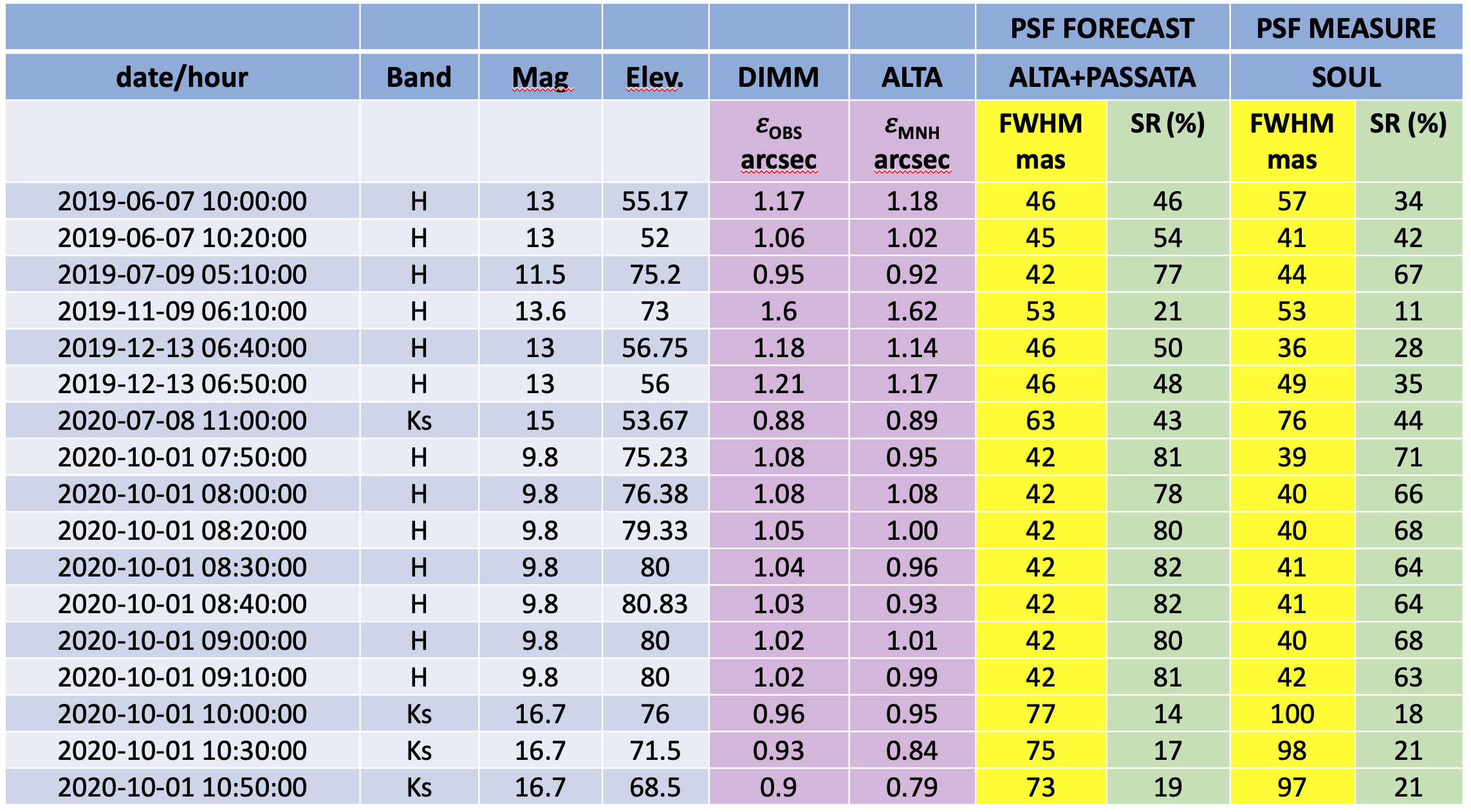}\\
\caption{Table reporting the full measurement dataset used for this work, together with the forecasts produced by our models. Dates are reported in UT. All measurement values are medians over 10-minutes intervals. Columns with the same color represent the values that must be compared. DIMM and ALTA columns represent the seeing measured and forecasted by the ALTA AR model respectively (on the zenith), which, projected on the line of sight, is used as an input for the ALTA+PASSATA PSF forecast. The SOUL column contains the corresponding FWHM and SR measurements.} 
\end{table}

In Table \ref{fig:TableRes} we report the full list of the data obtained in this study, as medians over 10-minutes intervals. As previously stated, by construction, the ALTA column values are extremely accurate with respect to the DIMM. This allows us to study in a clean way all the source of uncertainties between the AO model and the SOUL measurements (see columns ALTA+PASSATA and SOUL), which is the focus of our study at this stage.\\

This eventually allows us to highlight all the critical points in the evaluation of the measurement accuracy. Between the sources that are not perfectly modeled in PASSATA are the telescope structure vibrations that impact on the final PSF figures of merit. Another source which is currently included in the data and is not modeled by PASSATA are the NCPA. These are supposed to be corrected during science operations [\cite{esposito2020}], however must be taken into account for the validation of the PSF forecast. We are currently trying to evaluate the impact of NCPA and if it is possible to address the latter source during post-processing inside our model.\\

\begin{table}[htb]
 \begin{center}
 \begin{tabular}{c|cc}
 \hline
 \multicolumn{1}{c}{} & {\bf FWHM (mas)} & {\bf SR (\%)} \\
 \hline
  {\bf RMSE} & 10.9 & 12.5 \\
  {\bf BIAS} & -4.5 & 9.8 \\
  {\bf $\sigma$} & 9.9 & 7.8 \\
 \hline
 \end{tabular}
 \label{tab:stats}
 \end{center}
 \caption{Standard statistical indicators obtained by comparing the ALTA+PASSATA PSF forecasts and the SOUL measurements from Table \ref{fig:TableRes}, for FHWM and SR respectively.} 
\end{table}

We report in Table \ref{tab:stats} the average statistical indicators obtained from the comparison of the ALTA + PASSATA FWHM and SR forecast and the SOUL corresponding measurements, with an RMSE on FWHM and SR of 11 mas and 12.5\% respectively. The statistical sample is still too small to be quantitatively representative but this exercise allowed us to define the conditions of analysis. The next step is to enrich the statistical sample and to quantify the sources of uncertainty as those already mentioned (i.e. AO vibrations and NCPA). Part of the large BIAS experimented in this study, especially for the SR, may be explained by the NCPA which are not corrected in this SOUL dataset. This should eventually not affect the science data because in that case NCPA are typically corrected [\cite{esposito2020}], however in order to quantify the performance in post-processing we must be able to disentangle this contribution from the measurements. Looking at the perfomance on-sky of NCPA correction [\cite{esposito2020}], this may impact the SR by a factor greater that 15\%, thus our results in \ref{tab:stats} are indeed promising. To this we should add the uncertainty from the atmospheric model, which however is already well characterized and small (around 0.1 arcsec) [\cite{masciadri2020}].\\

\section{CONCLUSIONS}
\label{conclusion}

We present in this contribution a preliminary evaluation of the feasibility of a forecast system for the PSF figures of merit. Specifically we investigate if an atmospheric model of the OT coupled with an end-to-end AO modeling software (PASSATA), can be effective in forecasting the FWHM and SR for PSF measurements obtained with AO instruments. This study was initially applied to SOUL at LBT. We chose an initial simplified setup deciding to limit our study to the on-axis case, which gives no dependency of the PSF on the turbulence stratification, and thus our forecasts uses only the seeing value from the atmospheric forecasts as an input. Preliminary results of this study tells us that the uncertainty on the AO modeling/AO measurement is of the order of 12.5\% for the SR and 11 mas for the FWHM. By considering the typical impact of NCPA observed on-sky, our results can be considered promising. During the study we pointed out the critical problem of the assessment of the measurement accuracy, and the capability of PASSATA to model correctly the measurement variability. For the prosecution of our work, we intend to achieve the following steps:
\begin{itemize}
    \item to disentangle the contribution to the measurements, variability coming from noise sources such as AO vibrations and NCPA
    \item to enrich the sample of PSF measurements from SOUL in order to increase the statistics for the forecast validation,
    \item to implement SAXO/SPHERE inside PASSATA, allowing us to use the SAXO larger measurement sample,
    \item to include the uncertainty produced by the atmospheric model (which however is already known to be around 0.1 arcsec on 1-hour timescale),
    \item to consider also the off-axis PSF cases by including the full $C_N^2$ and wind stratification into the model.
\end{itemize}
Future key milestones of this project will be addressing the previous issues. In future studies we aim to investigate also the possibility to use intrinsically faster analytical models, such as the one described in Neichel et al. [\cite{neichel2008}], and possibly confront the accuracy obtainable with both approaches. The final goal of this work is the development of an operational forecast system for the PSF figures of merit, including FWHM, SR, but also Encircled Energy and contrast, could greatly benefit AO-assisted observation planning in flexible scheduling and also contribute to the runtime tuning of AO instrumentation.

\acknowledgments 
 
The study has been co-funded by the FCRF foundation through the 'Ricerca Scientifica e Tecnologica' action - N.45103. by the project ALTA Center (ENV001, ENV002) and by the European Union's Horizon 2020 research and innovation programme under grant agreement No 824135 (SOLARNET).

\bibliographystyle{spiebib} 

\end{document}